\documentclass{article}
\usepackage[utf8]{inputenc}
\usepackage{amsmath}
\usepackage{amsthm}
\usepackage{amssymb}
\usepackage{amsfonts}
\usepackage{esint}
\usepackage{physics}
\usepackage{graphics}
\usepackage{graphicx}
\usepackage{bm}
\usepackage[utf8]{inputenc}
\usepackage{fancyhdr}
\usepackage{float}
\usepackage{textcomp}
\usepackage[colorlinks = true,
            linkcolor = blue,
            urlcolor  = blue,
            citecolor = blue,
            anchorcolor = blue]{hyperref}
\usepackage{amsmath,amssymb}
\usepackage{physics}
\usepackage{wrapfig,lipsum}
\usepackage{mathtools}
\usepackage{longtable}
\usepackage{tikz}
\usetikzlibrary{shapes.geometric, arrows}
\usetikzlibrary{positioning}
\usepackage{epsfig}
\usepackage{graphicx}
\usepackage{color}
\usepackage{multirow}
\usepackage{listings}
\usepackage{authblk}
\usepackage{hyperref}
\usepackage{subcaption}
\usepackage{pgfplots}
\pgfplotsset{compat=1.16}
\pgfplotsset{every axis/.append style={tick label style={/pgf/number format/fixed},font=\scriptsize,ylabel near ticks,xlabel near ticks,grid=major}}
\newcommand{\RR}{\mathbb  R}

\newcommand{\comment}[1]{}

\numberwithin{equation}{section}

\DeclareMathOperator{\Isom}{\Isom}

\title{Classification of Diabetic Retinopathy Severity in Fundus Images with DenseNet121 and ResNet50}
\author[1,5]{Jonathan Zhang\thanks{jzhang.of.ny@gmail.com}}
\author[2,5]{Bowen Xie\thanks{iamawesome1234w@gmail.com}}
\author[3,5]{Xin Wu\thanks{xwu1.618@gmail.com}}

\author[4,5]{Rahul Ram\thanks{rahulrram04@gmail.com}}
\author[4,5]{David Liang\thanks{dliang7234@gmail.com}}

\affil[1]{Commack High School, Commack, NY 11725}
\affil[2]{Glenda Dawson High School, Pearland, TX 77581}
\affil[3]{Mira Loma High School, Carmichael, CA 95765}

\affil[4]{Ward Melville High School, East Setauket, NY 11733}
\affil[5]{Machine Learning, Camp Illumina}
\date{August 18, 2021}
\begin{document}

\maketitle

\begin{align*}
    \includegraphics[]{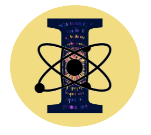}
\end{align*}

\centering{\Huge Illumina Learning \par}

\newpage

\tableofcontents
\newpage
\lhead{\includegraphics[scale=0.25]{logo.png}}
\chead{}
\large

\begin{abstract}
    In this work, deep learning algorithms are used to classify fundus images in terms of diabetic retinopathy severity. Six different combinations of two model architectures, the Dense Convolutional Network-121 and the Residual Neural Network-50 and three image types, RGB, Green, and High Contrast, were tested to find the highest performing combination. We achieved an average validation loss of 0.17 and a max validation accuracy of 85 percent. By testing out multiple combinations, certain combinations of parameters performed better than others, though minimal variance was found overall. Green filtration was shown to perform the poorest, while amplified contrast appeared to have a negligible effect in comparison to RGB analysis. ResNet50 proved to be less of a robust model as opposed to DenseNet121.
 
\end{abstract}

\section{Introduction}

 Diabetic Retinopathy (DR), is a form of vision impairment caused by high blood sugar. The high amount of blood sugar damages the blood vessels in the retina, and causes blood leakage in the eye. This can lead to vision loss and if untreated, can lead to blindness. DR is the leading cause of vision impairment in working-age adults, and it is estimated that 200 million people in the world will suffer from DR by 2050. So far, the safest and most accurate way to check for DR has been DR screenings, where human assessors check the retina to make sure it is not damaged. Unfortunately, this method is not widely performed because of a lack of qualified assessors, which led us to seek a machine learning model to solve this issue. Most of our models perform well, with an average testing/validation accuracy of 80\%, which is impressive considering the amount of time we had to experiment.

\newpage
\section{Methodology}

\subsection{Overview of Proposed Study}

To solve the problem of a lack of human assessors for DR, we compiled a neural network in Python to classify DR severity levels in different fundus images. We obtained our data from kaggle.com, where the dataset was hosted \hyperlink{1}{[1]}. It was hosted by the Asia Pacific Tele-Ophthalmology Society (APTOS) as a competition to increase awareness for Diabetic Retinopathy. For our experiment, we tested 6 different combinations of CNN model architectures and image types to find the combination with the best validation accuracy and loss. The table below shows all of the combinations that we tested on our fundus image dataset.

\vspace{5mm}

\begin{table}[hbt!]
    
    \centering
    \hspace*{-1cm}\begin{tabular}{|c|c|c|c|}
     
     \hline
      & \textbf{Original (RGB)} & \textbf{Green} & \textbf{High Contrast (HC)} \\ 
     \hline
     \textbf{DenseNet121} & DenseNet121 + RGB & DenseNet121 + Green & DenseNet121 + HC \\ 
     \hline
     \textbf{ResNet50} & ResNet50 + RGB & ResNet50 + Green & ResNet50 + HC \\ 
     \hline
    
    \end{tabular}\hspace*{-1cm}
    \caption{Combinations of model architectures and image types.}
\end{table}

\vspace{5mm}

We followed the general steps below to create our model and classify fundus images.

\vspace{5mm}

\tikzstyle{block} = [rectangle, draw,  
    text width=4.5em, text centered, rounded corners, minimum height=4em] 
\tikzstyle{line} = [draw, -latex']
\begin{tikzpicture}[node distance = 3cm, auto] 
    \node [block] (init) {Import images};  
    \node [block, right of= init](ReadA){Import diagnosis data};  
    \node [block, right of= ReadA](ReadB){Convert images/ diagnoses into array};  
    \node [block, below of = ReadB](Sum){Set aside 20\% of data for testing};  
    \node [block, left of = Sum](P){Create model};  
    \node [block, left of = P](Out){ Fit model to data};  
    \path [line] (init) -- (ReadA);   
    \path [line] (ReadA) -- (ReadB);   
    \path [line] (ReadB) -- (Sum);   
    \path [line] (Sum) -- (P);   
    \path [line] (P) -- (Out);   
      
\end{tikzpicture}  

\subsection{Preprocessing Stage}
Before training our models with the 3662 RGB fundus images from the Kaggle dataset \hyperlink{1}{[1]}, we had to preprocess our images to help the model process the images more easily. The images were in .png format and had varying resolutions. We ended up choosing 3 different image types to test, the original RGB version, a green-only filter, and a high contrast filter, as they have been shown to perform well with image classification \hyperlink{9}{[9]}. The green and high contrast images were created by altering the arrays of the images. For the green-only filter, we replace the values of the red and blue columns of the images array with zeroes, creating an image with only green values. For the high contrast filter, we split the image into 3 layers, red, green, and blue. Then, we equalize the histogram of each layer of the image, and merge them together to create the high contrast image. Below are 3 of the same fundus image under different filters.

\hspace*{-1cm}\begin{figure}[htp]

\centering
\includegraphics[width=.3\textwidth]{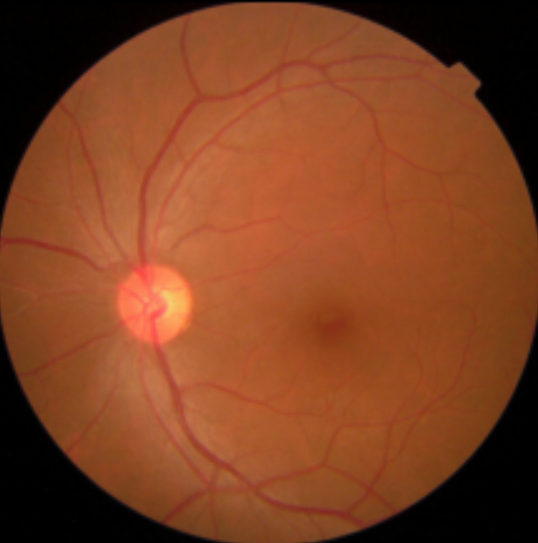}\hfill
\includegraphics[width=.3\textwidth]{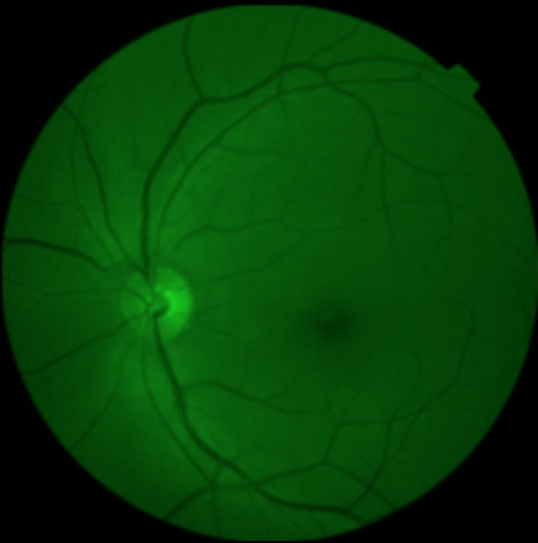}\hfill
\includegraphics[width=.3\textwidth]{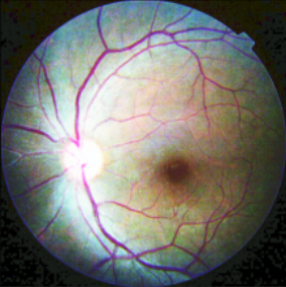}

\caption{The three image types we tested in our project.}
\label{fig:figure3}

\end{figure}\hspace*{-1cm}

We also had to resize all of our images to a 244x244 resolution to make the images smaller and easier to process, and to match the default input dimensions for DenseNet121 and ResNet50, the architectures we used to compile our models.

\subsection{Network Architecture}

In our research, we used neural networks to help us with our machine learning algorithms. Neural networks are a subset of machine learning and consists of neurons in the form of 
\[ a =f(z)=f(\sum_{i=1}^{m}x_{i}w_{i}+w_0)=f(\bm{w}^{T}\bm{x}+w_0) \]

where $a$ is the output, $z$ is the weighted sum, $f$ is the activation function, $\bm{w}$ is the weight, and $\bm{x} \in \RR^m$ is the $m$-dimensional input vector. We can train a neuron by optimizing its loss function through gradient descent, which we chose to be \textit{binary cross-entropy}, defined by

\[ L(t, \hat{y}) =-\frac{1}{t}\sum_{i=1}^{t}y_{i}\log \hat y_{i}+(1-y_{i})\log (1-\hat y_{i}) \]
where $t$ is the output size, given by $\frac{W-K+2P}{S}+1$, $W$ is the input size which is initially $224$, $K$ is the kernel size, which varies from $3$ to $7$ from convolution layer to layer, and $P$ is the image padding, which we set to $10$ so the original $244$x$244$ images fit our DenseNet121 and ResNet50 implementation requirements. $\bm{y}$ represents a vector of expected outcome whereas $\bm{\hat{y}}$ represents a vector of predictions.

Due to its memory advantages with respect to the positive classes we defined, we used the rectified linear unit (ReLU) activation function defined as 

\[ R(z) := \begin{cases} 0 & \text{if } z<0 \\ z &\text{otherwise} \end{cases}=\max(0,z) \]

The convolutional neural network (CNN), commonly used to analyze images, was utilized to process fundus images used in DR screenings. The main concept behind CNNs is convolution, which is a math operation that lets us find a pattern in a portion of the image. To perform it, we multiply each number in an image tile by the corresponding number in the kernel, which is the pattern we are looking for. In the end, we add them together, and get an output. We do this for each pixel in an image, which allows us to find where the object in the image. The equation for convolution is defined as

\[G[m,n] = \sum_{j}\sum_{k}h[j,k]f[m-j,n-k]\]

where $f$ is the input image, $h$ is the kernel, $m$ is the index of the row, and $n$ is the index of the column.

The 2 types of model architectures we used were DenseNet121 and ResNet50. In a DenseNet121 architecture, each layer is connected to every other layer, while layers only connect to the next layer in traditional CNNs \hyperlink{8}{[8]}. This allows for fewer parameters, making DenseNet more efficient than traditional CNNs. In a ResNet50 architecture, shortcut connections are introduced that skip one or more layers, which allows the network to be trained much deeper than normal \hyperlink{5}{[5]}. 

\subsection{Model Compilation and Training Process}
In our code, we used Tensorflow and Keras to compile our Sequential models, along with DenseNet121 and ResNet50. Many hyperparameters needed to be changed for our model, including the learning rate, optimizer and number of epochs. The learning rate is an optimization algorithm that controls how much the model is changed in response to a prediction error \hyperlink{3}{[3]}. The optimizer we used was Adam, introduced in 2015 as an alternative to the stochastic gradient descent procedure (SGD). We chose Adam as the optimizer because it is memory-efficient and easy to implement. For our number of epochs, we chose 15 because it gave all of our combinations a chance to reach its highest validation accuracy. We also chose a batch size of 32, as it is an optimal batch size for image classification models.

\section{Results}

Below are the plots for our model statistics. 

\hspace*{-1cm}\begin{figure}[htp]

    \centering
    \includegraphics[width=6cm]{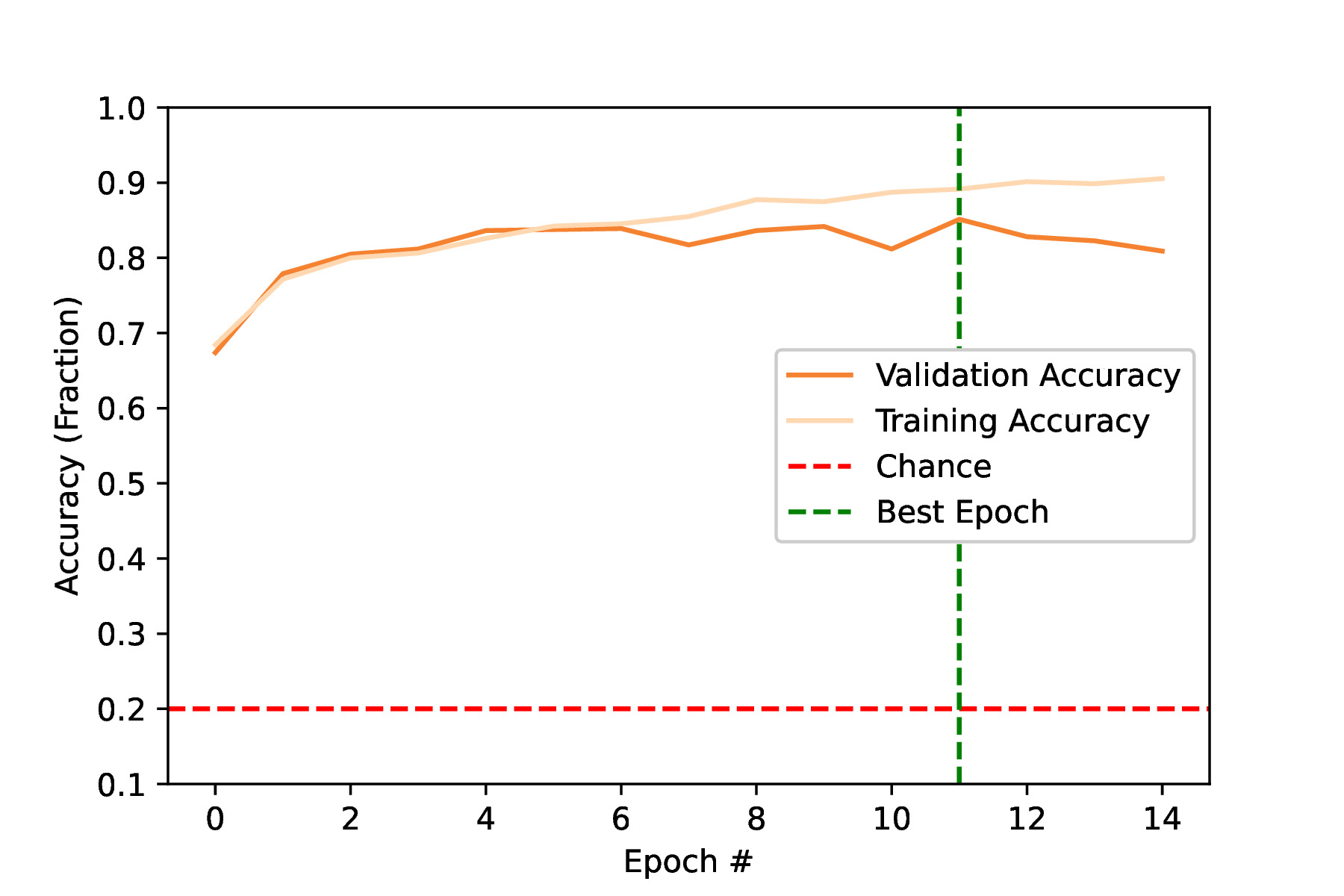}
    \includegraphics[width=6cm]{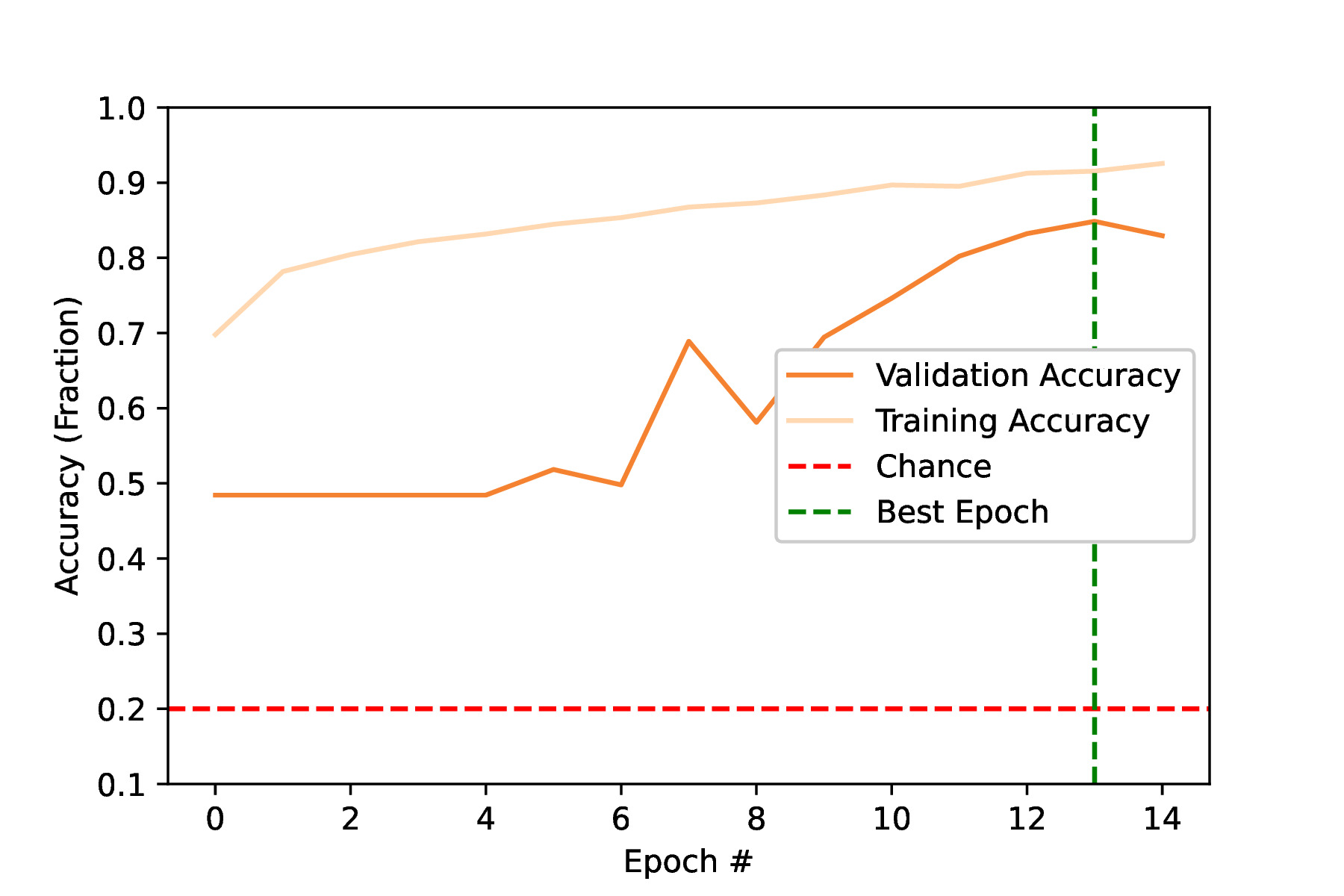}
    \caption{(\textbf{Left}) DenseNet121 + RGB Accuracy (\textbf{Right}) ResNet50 + RGB Accuracy}

\end{figure}\hspace*{-1cm}

\hspace*{-1cm}\begin{figure}[htp]

    \centering
    \includegraphics[width=6cm]{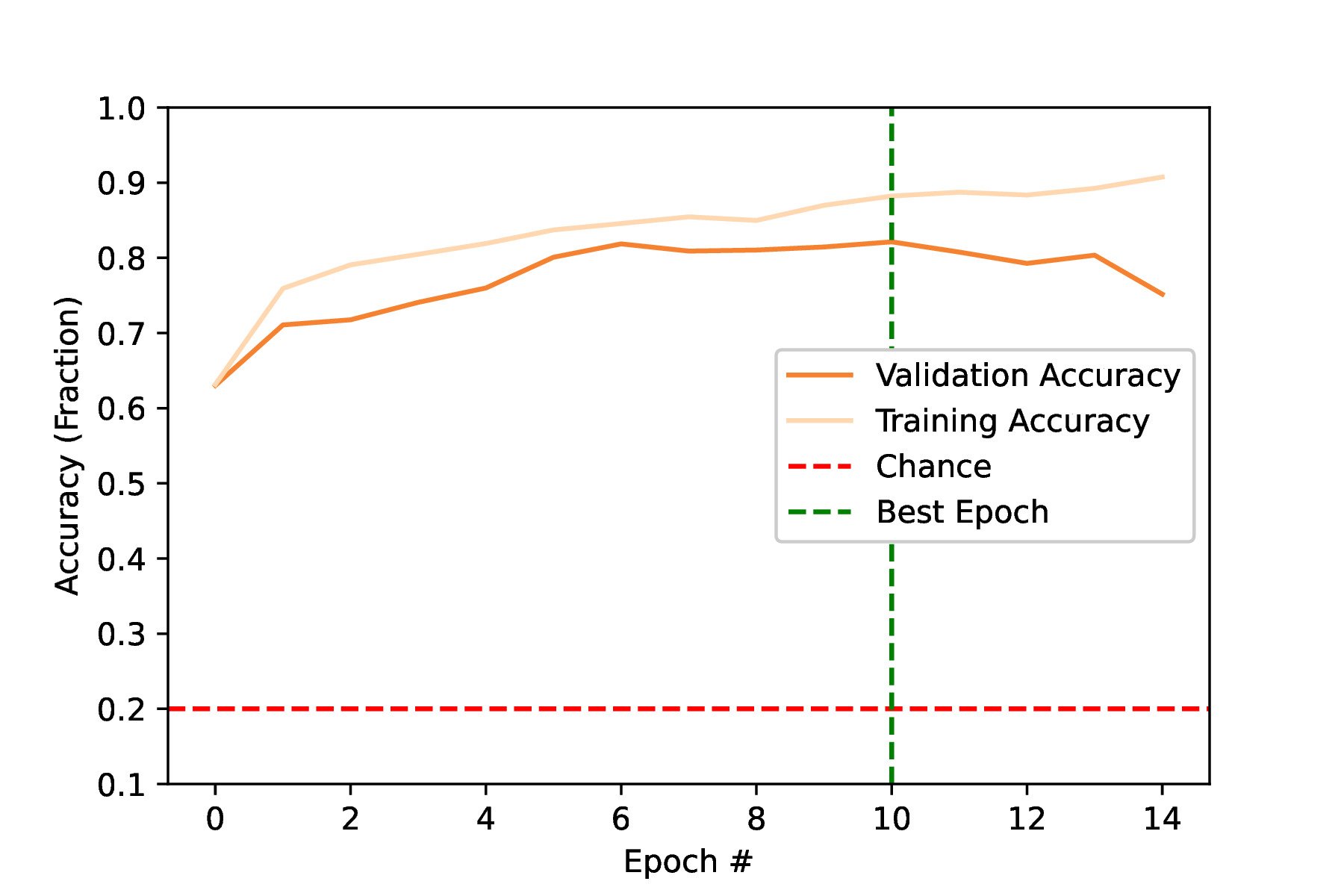}
    \includegraphics[width=6cm]{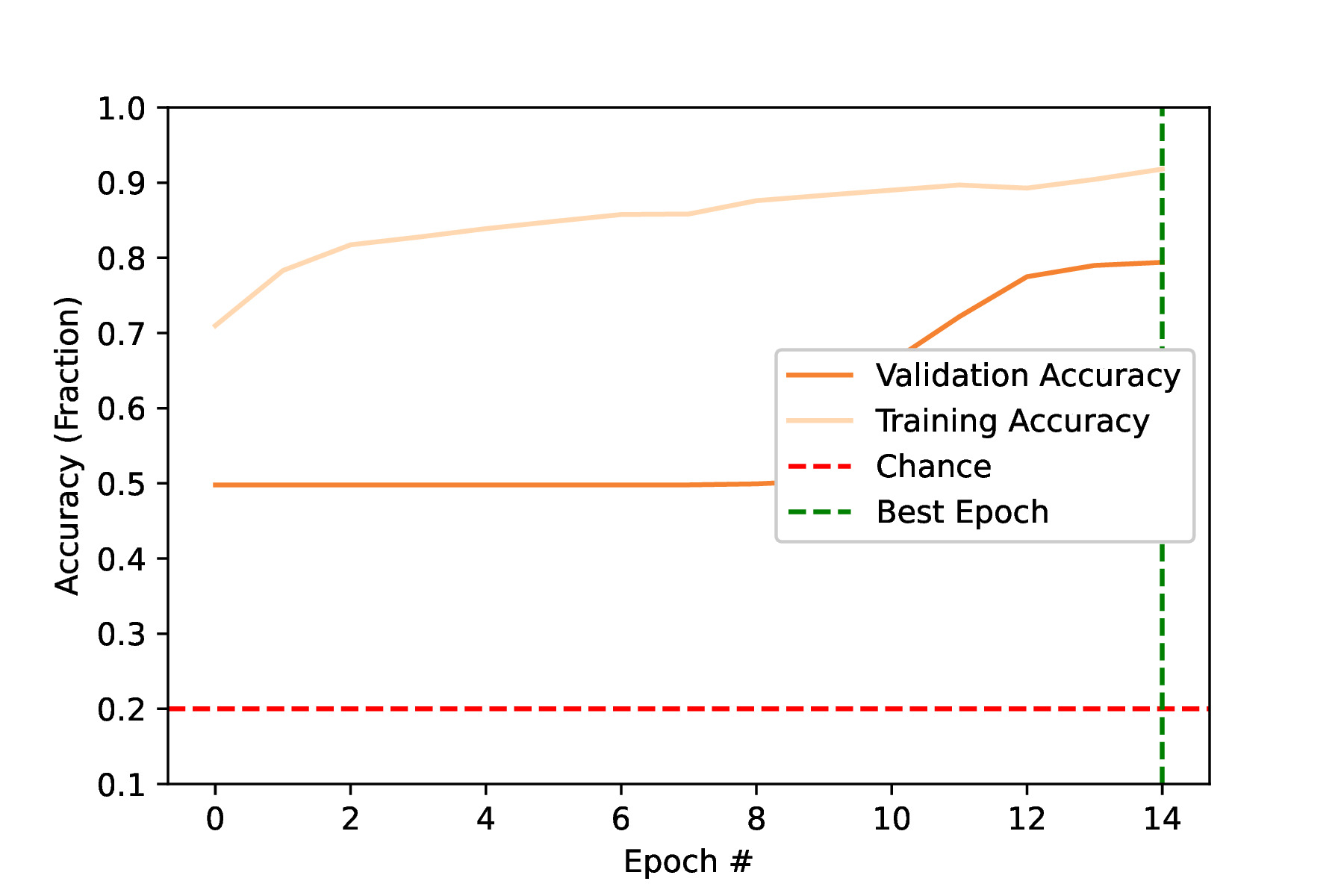}
    \caption{(\textbf{Left}) DenseNet121 + Green Accuracy (\textbf{Right}) ResNet50 + Green Accuracy}

\end{figure}\hspace*{-1cm}

\hspace*{-1cm}\begin{figure}[htp]

    \centering
    \includegraphics[width=6cm]{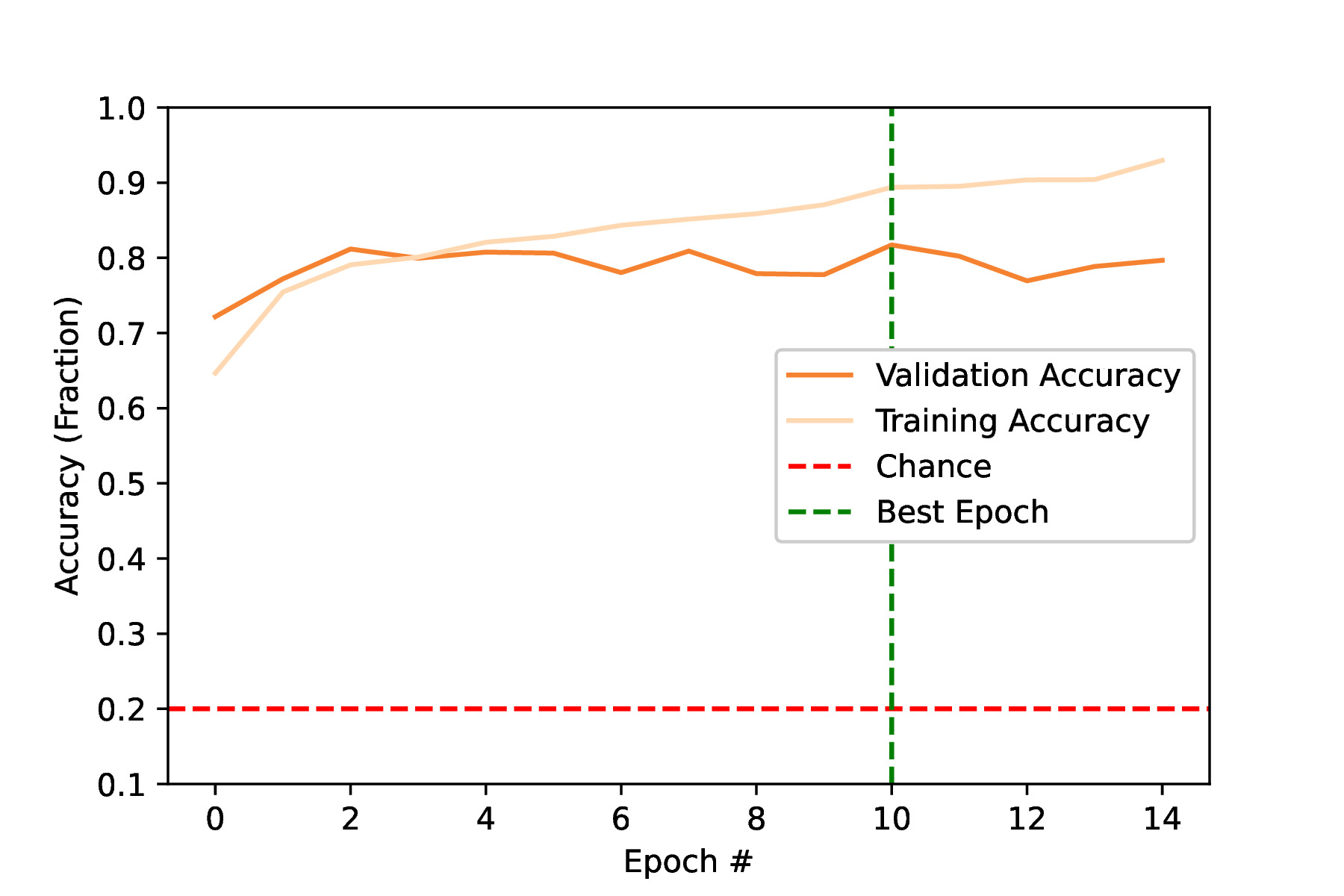}
    \includegraphics[width=6cm]{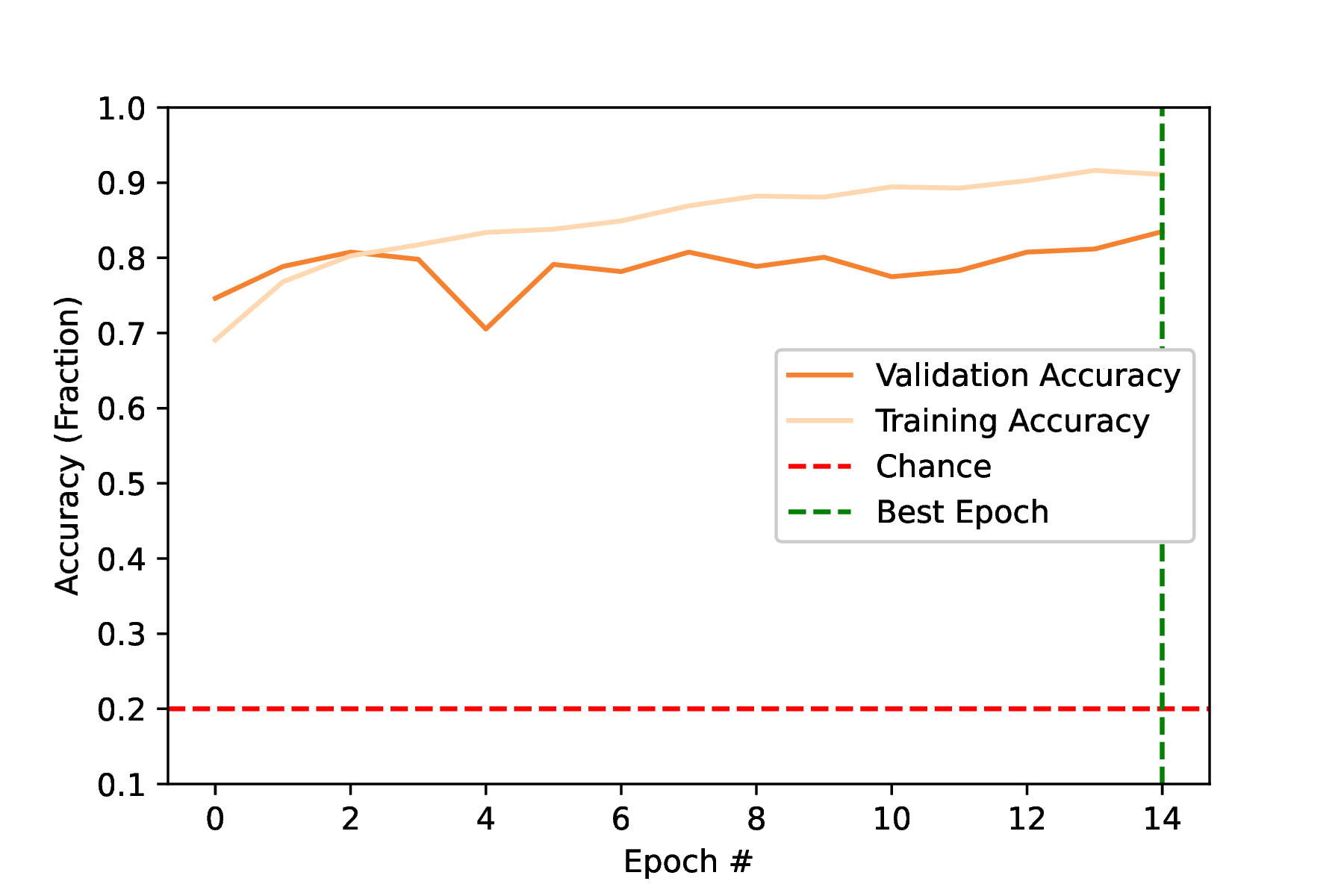}
    \caption{(\textbf{Left}) DenseNet121 + High Contrast Accuracy (\textbf{Right}) ResNet50 + High Contrast Accuracy}

\end{figure}\hspace*{-1cm}

\hspace*{-1cm}\begin{figure}[htp]

    \centering
    \includegraphics[width=6cm]{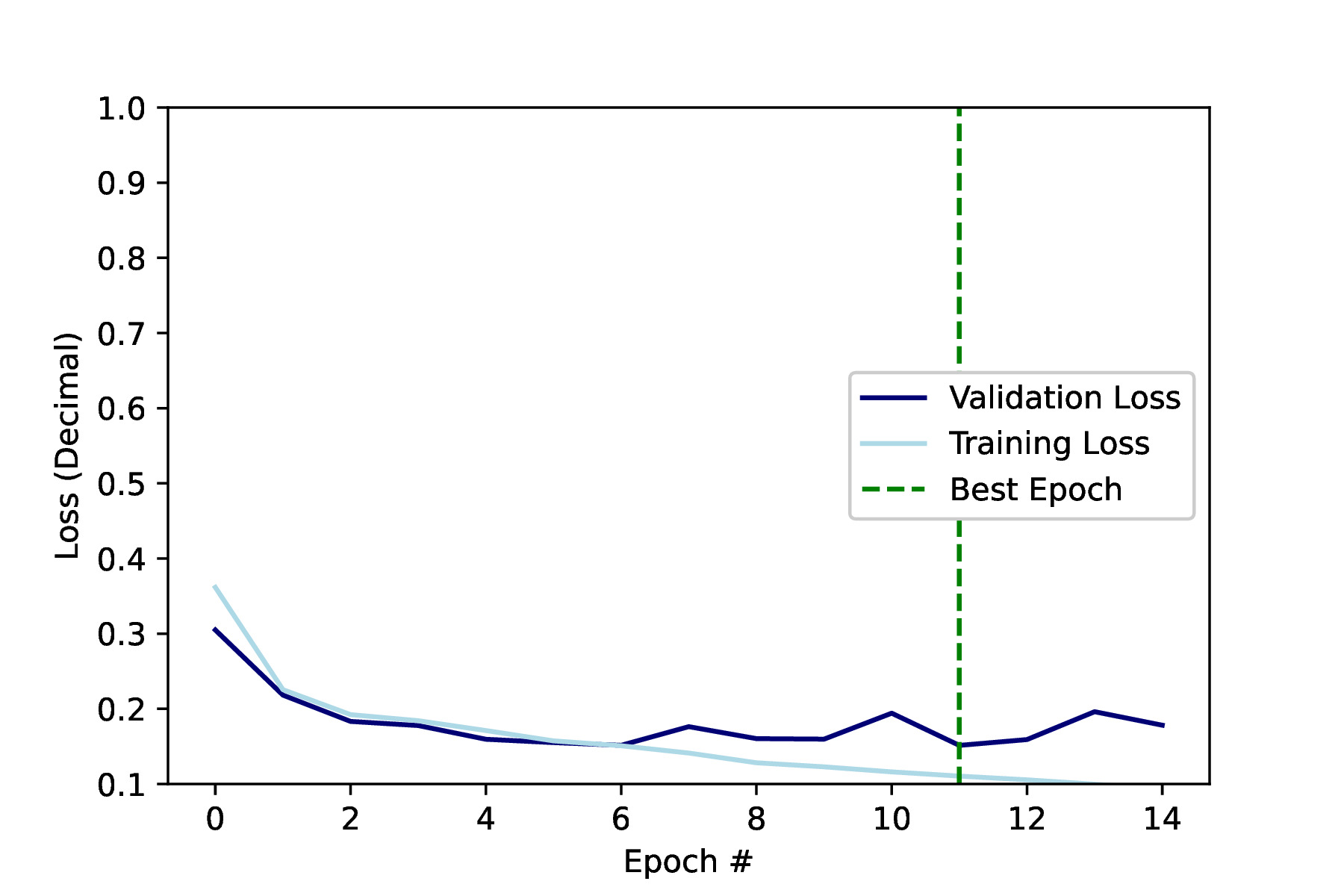}
    \includegraphics[width=6cm]{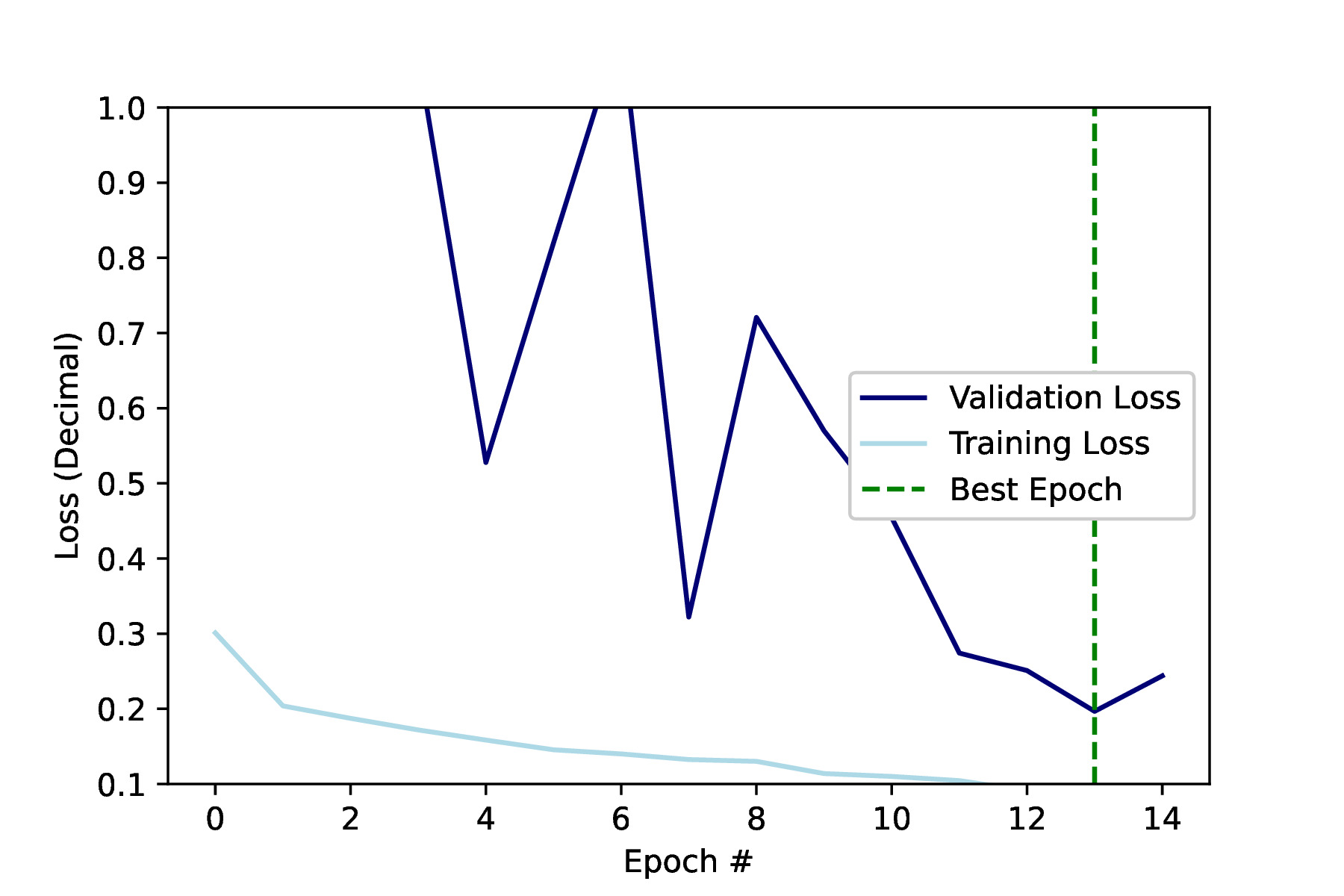}
    \caption{(\textbf{Left}) DenseNet121 + RGB Loss (\textbf{Right}) ResNet50 + RGB Loss}

\end{figure}\hspace*{-1cm}

\hspace*{-1cm}\begin{figure}[htp]

    \centering
    \includegraphics[width=6cm]{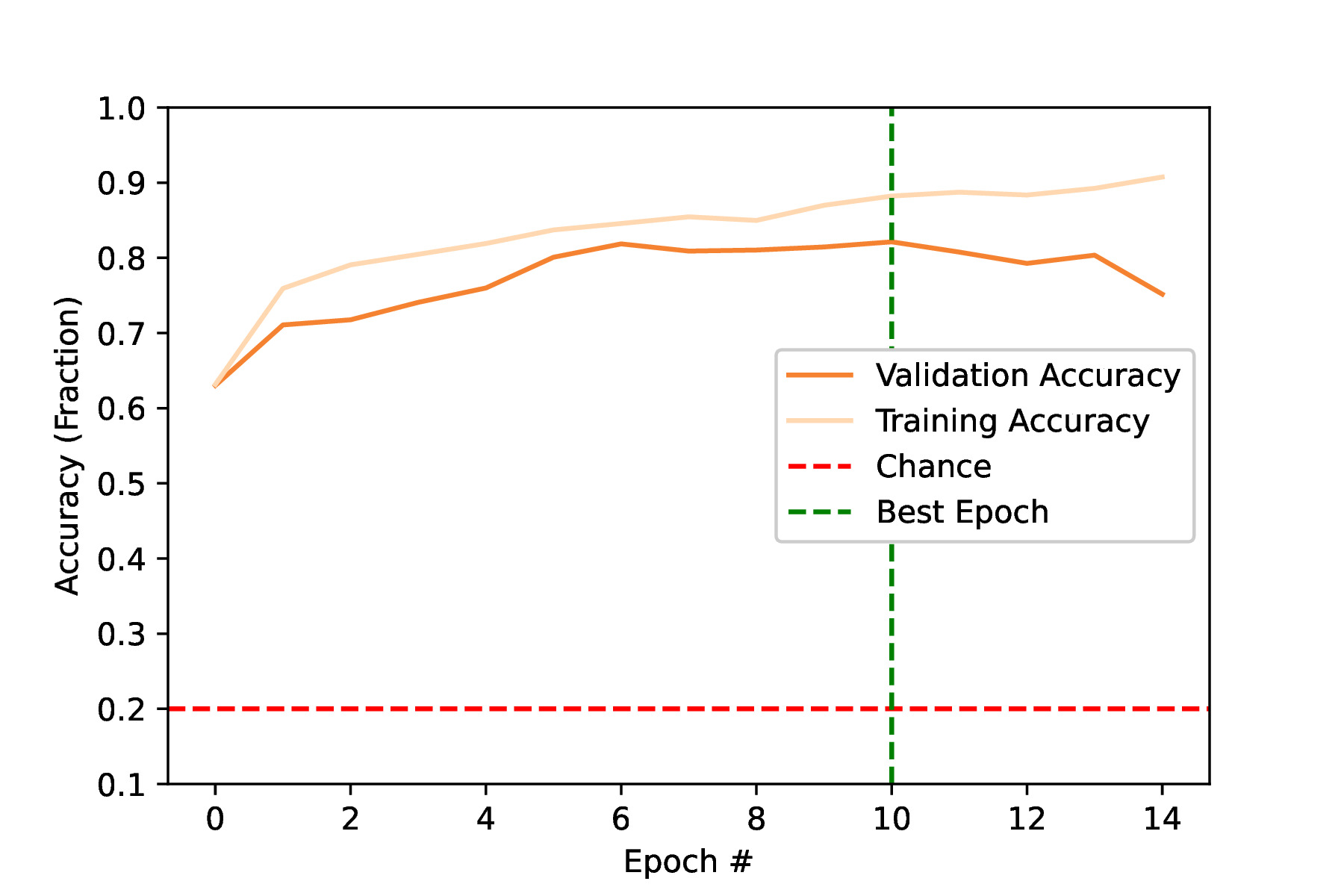}
    \includegraphics[width=6cm]{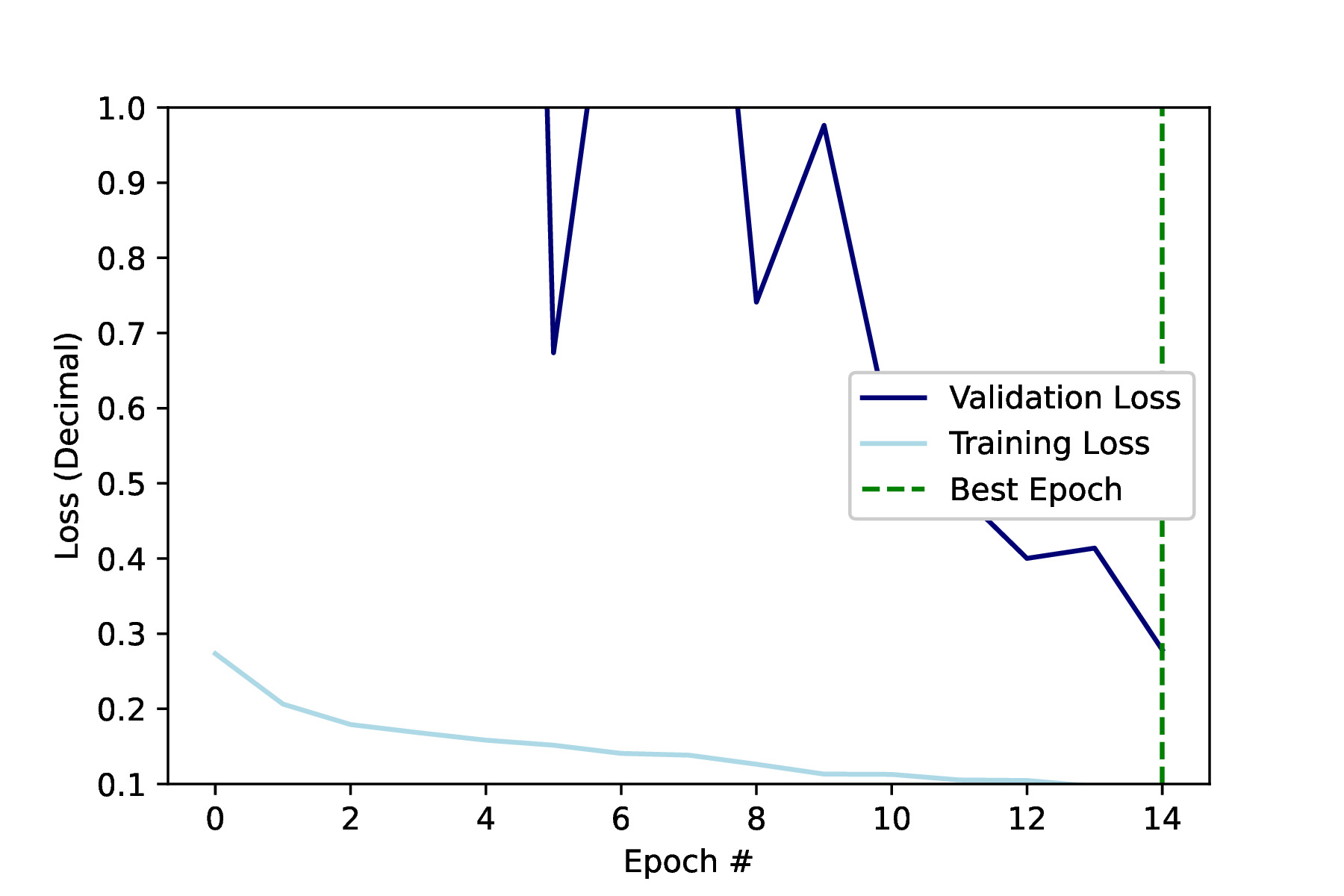}
    \caption{(\textbf{Left}) DenseNet121 + Green Loss (\textbf{Right}) ResNet50 + Green Loss}

\end{figure}\hspace*{-1cm}

\hspace*{-1cm}\begin{figure}[htp]

    \centering
    \includegraphics[width=6cm]{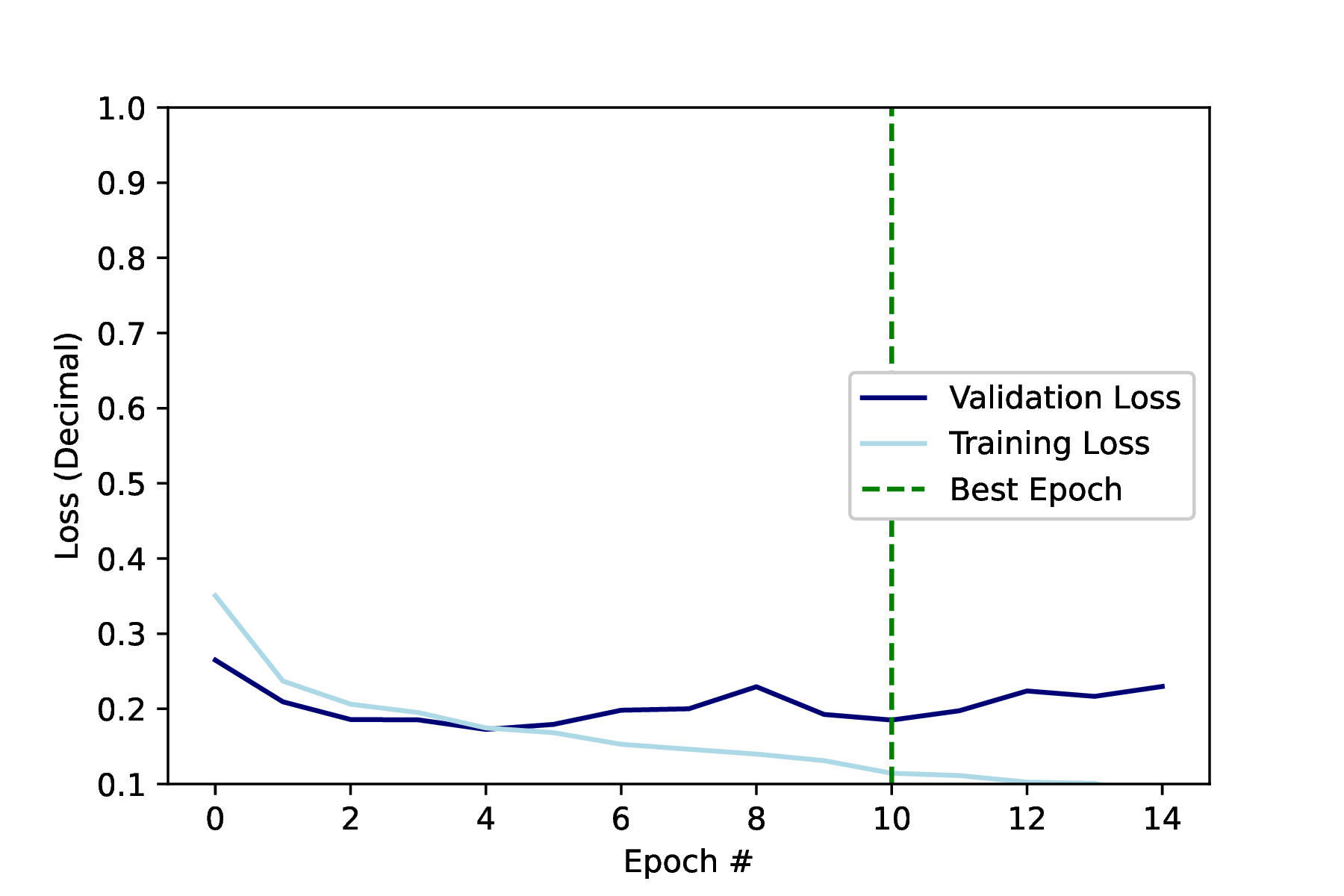}
    \includegraphics[width=6cm]{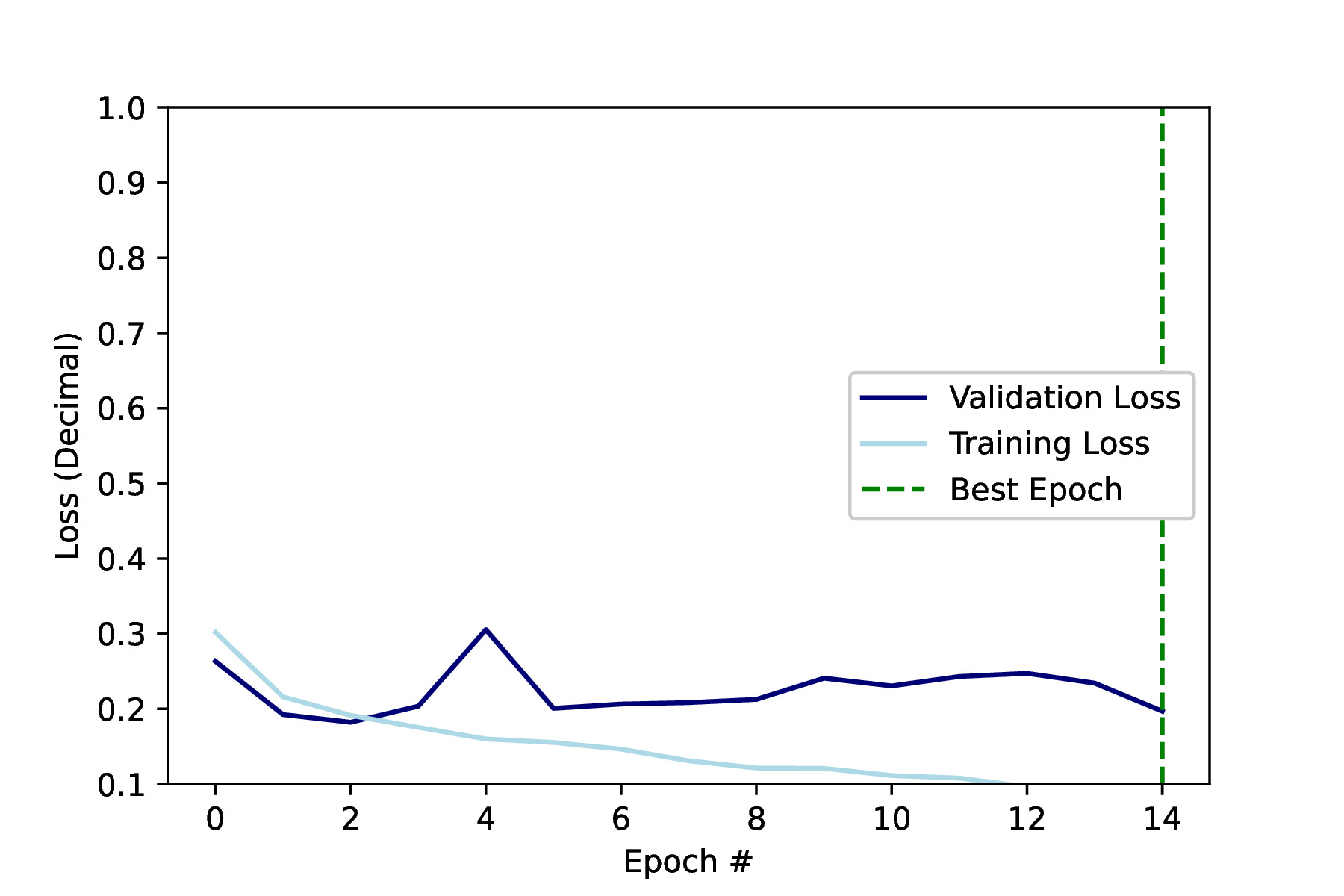}
    \caption{(\textbf{Left}) DenseNet121 + High Contrast Loss (\textbf{Right}) ResNet50 + High Contrast Loss}

\end{figure}\hspace*{-1cm}

\begin{table}[hbt!]
    
\centering
\hspace*{-2cm}\begin{tabular}{|c|c|c|c|}
     
    \hline
    & \textbf{DN121 + RGB} & \textbf{DN121 + Green} & \textbf{DN121 + HC} \\
    \hline
    & \textbf{RN50 + RGB} & \textbf{RN50 + Green} & \textbf{RN50 + HC} \\ 
    \hline
    \textbf{Average Validation Accuracy* (\%) } & $82.7$\% & $79.2$\% & $78.9$\% \\
    \hline
    & $83.7$\% & $79.4$\% & $83.4$\% \\ 
    \hline
    \textbf{Average Validation Loss* ($0 \leq L \leq 1$)} & $0.17$ & $0.21$ & $0.21$ \\
    \hline
    & $0.22$ & $0.28$ & $0.20$ \\ 
    \hline
    
    \end{tabular}\hspace*{-1cm}
    \caption{Table of results for our experiment. \\ *After the best epoch (higher accuracies and lower losses are favorable).}
\end{table}

\newpage
\section{Discussion}


\subsection{Result Analysis}
From the table of results, we can see that there is no significant advantage that one combination has over the others, but only slight advantages in different categories. ResNet50 + RGB had the highest average validation accuracy, but DenseNet121 + RGB had the lowest average validation loss. As a result, we come to the conclusion that we simply do not have enough data to get a firm answer as to which combination performs the best. However, we did acheive fairly high results throughout most of our slides, but these numbers can always be improved.

\subsection{Future Works}
In the future, it is possible for us to add more datasets to our collection to increase our accuracy and decrease our loss. It is also possible for us to test out different architectures, as we only tried two, DenseNet121 and ResNet50. We could try out AlexNet or VGG in the future, as they are both well-respected model architectures in the field of image classification. Also, it may be possible for us to use our model to classify for other diseases like glaucoma and age-related macular degeneration (AMD) in the future. These are both eye diseases related to vision loss and blindness in adults. Most importantly, implementing feature extraction into these architectures could make our model much stronger by quantitatively analyzing the diabetic retinopathy image data.

\newpage

\section{Appendix}

\subsection{Diagrams and Figures}

\subsubsection{Activation Functions}

Below is the graph for the rectified linear unit function, $R(z) := \max(0,z)$, which is the primary activation function that was applied through each layer in both DenseNet121 and ResNet50.

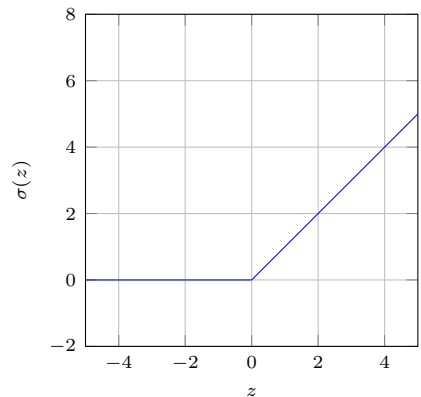
\begin{figure}[h]
    \centering
    \begin{tikzpicture}[declare function={sigma(\x)=1/(1+exp(-\x));}]
        \begin{axis}[width=6cm,height=6cm,ylabel=$\sigma(z)$,xlabel=$z$,ymin=-2,ymax=8,xmin=-5,xmax=5]
            \addplot[blue,smooth, domain=-5:0] {0};
            \addplot[blue,smooth, domain=0:5] {x};
        \end{axis}
    \end{tikzpicture}
    \captionsetup{width=0.5\textwidth}
    \caption{Graph of rectified linear unit activation function, $R(z) := \max(0,z)$.}
\end{figure}

We additionally applied Softmax at the end junction of our DenseNet121 network, defined as the vector-valued function $\bm{\sigma}$ with the following calculable indices:

\[ \sigma(\bm{z})_{i}=\frac{e^{z_{i}}}{\sum_{j=1}^{K} e^{z_{j}}} \]

where $\bm{\sigma}$ is the Softmax function, $\bm{z}$ is the input vector from the penultimate layer of the network, and $K = 5 \; (\{0,1,2,3,4\})$ is the number of classes in our multi-class classifier. We can additionally define the above function in a more compact sense defining the exponential of a vector:

\begin{align}e^{\bm{v}}&:= \sum_{n = 0}^{\infty} \frac{\bm{v}^n}{n!}  = 1+\bm{v}+\frac{\Vert\bm{v}\Vert^2}{2!}+\frac{\Vert\bm{v}\Vert^2}{3!}\bm{v}+\cdots
\\&=\begin{cases}1 & \text{if }\bm{v}=\bm{0}\\1+\Vert\bm{v}\Vert\frac{\bm{v}}{\Vert\bm{v}\Vert}+\frac{\Vert\bm{v}\Vert^2}{2!}+\frac{\Vert\bm{v}\Vert^3}{3!}\frac{\bm{v}}{\Vert\bm{v}\Vert}+\cdots & \text{otherwise}\end{cases}
\\&=\begin{cases}1 & \text{if }\bm{v}=\bm{0}\\\cosh\left(\Vert\bm{v}\Vert\right)+\frac{\bm{v}}{\Vert\bm{v}\Vert}\sinh\left(\Vert\bm{v}\Vert\right) & \text{otherwise}\end{cases}
\end{align}

\subsubsection{Architecture Diagrams}

\hspace*{-2cm}\begin{figure}[hbt]
    \centering
    \includegraphics[scale=1, width=12cm]{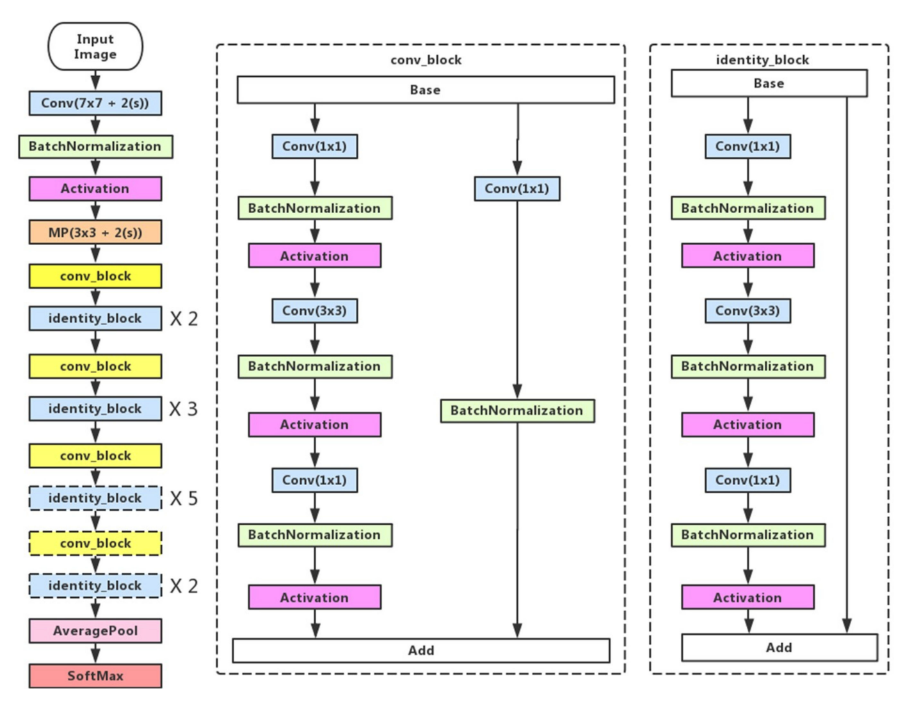}
    \caption{(\textbf{Left}) ResNet50 architecture. Blocks with dotted line represents modules that might be
removed in our experiments. (\textbf{Middle}) Convolution block which changes the dimension of the input.
(\textbf{Right}) Identity block which will not change the dimension of the input.}
    
\end{figure}\hspace*{-2cm}

\hspace*{-2cm}\begin{figure}[hbt!]
    \centering
    \includegraphics[scale=1, width=12cm]{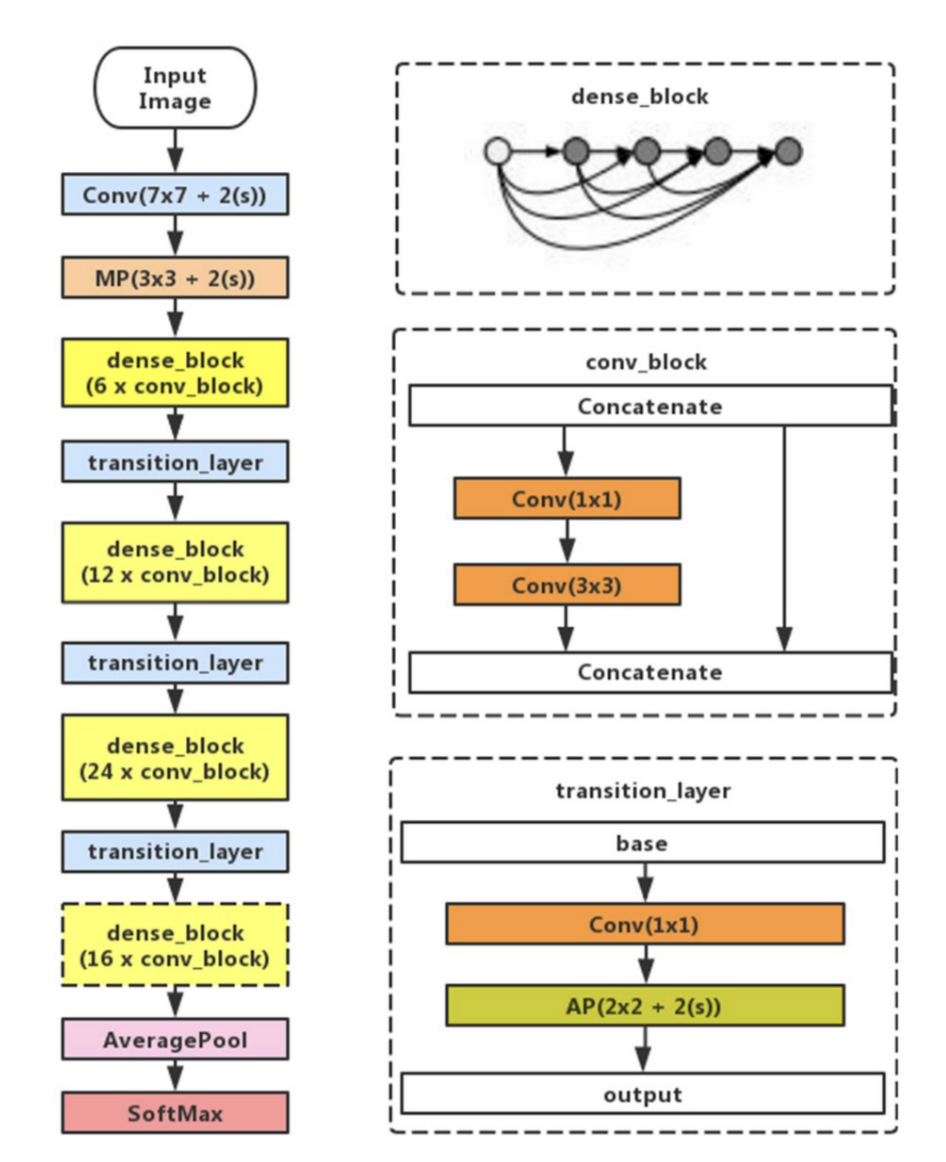}
    \caption{(\textbf{Left}) DenseNet121 architecture. (\textbf{Right}) Dense\_Block, conv\_block, and transition\_layer}
    
\end{figure}\hspace*{-2cm}

\newpage
\subsection{Code}
\vspace{5mm}

Full Code: 
\\\href{https://github.com/JZhang-305/Diabetic-Retinopathy-Classifier}{Diabetic Retinopathy Classifier Repository}

\vspace{5mm}

\begin{lstlisting}[language = Python]
# Function to compile our model
def compile_model(input_shape, DenseNet=True, ResNet=False):
    if DenseNet:
        model_type = DenseNet121(weights='../input/densenet/densenet121_weights_tf_dim_ordering_tf_kernels_notop.h5', include_top=False, input_shape=input_shape)
    elif ResNet:
        model_type = ResNet50(weights='imagenet', include_top=False, input_shape=input_shape)
        
    model = Sequential()
    model.add(model_type)
    model.add(GlobalAveragePooling2D())
    model.add(Dropout(0.5))
    model.add(Dense(5, activation='softmax'))

    model.compile(optimizer=Adam(lr=0.0001), loss='binary_crossentropy', metrics=['accuracy'])

    datagen = ImageDataGenerator(
        zoom_range=0.2,
        fill_mode='constant',
        cval=0.,
        horizontal_flip=True,
        vertical_flip=True)

    model.summary()
    
    return model
    

# Fit model to data and plot stats
history = model.fit(datagen.flow(x_train, y_train, batch_size=32), validation_data=(x_test, y_test), steps_per_epoch=(len(x_train)/32), epochs=15)
\end{lstlisting}
\section{Acknowledgments}
We would like to thank our mentors Rahul Ram and David Liang for their continued support and guidance throughout our research project. We would also like to thank Camp Illumina and its co-founders Amber Luo and Amy Liu for this opportunity to connect with others through research as a common field of passion. Finally, we would like to thank all of the RSI alumnus who acted as nobodies and gave us feedback during our posterless and final symposium presentations.

\newpage

\section{References}

\begin{flushleft}

\begin{small}

\hypertarget{1}{[1]} APTOS. (2019, June). APTOS 2019 Blindness Detection, Version 1. Retrieved August 13, 2021 from \url{https://www.kaggle.com/c/aptos2019-blindness-detection/data}.
\vspace{5mm}

[2] Arora, A. (2020, August 2). Densenet architecture explained with pytorch implementation from torchvision. Committed towards better future. \url{https://amaarora.github.io/2020/08/02/densenets.html\#densenet-architecture-introduction}. 

\vspace{5mm}

\hypertarget{3}{[3]} Brownlee, J. (2020, September 11). Understand the impact of learning rate on neural network performance. Machine Learning Mastery. \url{https://machinelearningmastery.com/understand-the-dynamics-of-learning-rate-on-deep-learning-neural-networks/}. 

\vspace{5mm}

[4] Dai, L., Wu, L., Li, H., Cai, C., Wu, Q., Kong, H., Liu, R., Wang, X., Hou, X., Liu, Y., Long, X., Wen, Y., Lu, L., Shen, Y., Chen, Y., Shen, D., Yang, X., Zou, H., Sheng, B., Jia, W. (2021, May 28). A deep learning system for detecting diabetic retinopathy across the disease spectrum. Nature News. \url{https://www.nature.com/articles/s41467-021-23458-5}. 

\vspace{5mm}

\hypertarget{5}{[5]} Feng, V. (2017, July 17). An overview of resnet and its variants. Medium. \url{https://towardsdatascience.com/an-overview-of-resnet-and-its-variants-5281e2f56035}. 

\vspace{5mm}

[6] Gheisari, S., Shariflou, S., Phu, J., Kennedy, P. J., Agar, A., Kalloniatis, M., Golzan, S. M. (2021, January 21). A combined convolutional and recurrent neural network for enhanced glaucoma detection. Nature News. \url{https://www.nature.com/articles/s41598-021-81554-4}. 

\vspace{5mm}

[7] Ji, Q., Huang, J., He, W., Sun, Y. (2019). Optimized Deep Convolutional Neural Networks for Identification of Macular Diseases from Optical Coherence Tomography Images. Algorithms, 12(3), 51. \url{doi:10.3390/a12030051}

\vspace{5mm}

\hypertarget{8}{[8]} Ruiz, P. (2018, October 18). Understanding and Visualizing densenets. Medium. \url{https://towardsdatascience.com/understanding-and-visualizing-densenets-7f688092391a}. 

\vspace{5mm}

\hypertarget{9}{[9]} Sisodia D. S, Nair S, Khobragade P. Diabetic Retinal Fundus Images: Preprocessing and Feature Extraction for Early Detection of Diabetic Retinopathy. Biomed Pharmacol J 2017;10(2).

\end{small}

\end{flushleft}

\end{document}